%% file: paper.tex
\documentclass[10pt,journal,compsoc]{IEEEtran}

\usepackage{url}
\usepackage{amsmath,amsfonts}
\usepackage[linesnumbered,ruled]{algorithm2e}
\usepackage{graphicx}
\usepackage{textcomp}
\usepackage{xcolor}
\usepackage{ tipa }
\usepackage{tabularx}
\usepackage{url}
\usepackage{color}
\usepackage{soul}
\usepackage{caption}
\usepackage{subcaption}
\usepackage{mwe}
\usepackage{enumerate}
\usepackage{pgfplots, pgfplotstable}
\usepackage{tikz}
\usepackage[utf8]{inputenc}
\usepackage[english]{babel}
\usepackage{amsmath}
\usepackage{amsthm}
\usepackage{mathtools}
\usepackage{csquotes}
\usepackage{blkarray, bigstrut}
\usepackage{gauss}
\usepackage[tikz]{bclogo}
\usepackage[framemethod=tikz]{mdframed}
\usepackage{lipsum}
\usepackage{booktabs}

\usepackage[most]{tcolorbox}

\tcbset{textmarker/.style={
        enhanced,
        parbox=true,boxrule=0mm,boxsep=1mm,arc=0mm,
        outer arc=1mm,left=2mm,right=0mm,top=7pt,bottom=7pt,
        toptitle=1mm,bottomtitle=1mm,oversize}}

\newtcolorbox{hintBox}{textmarker,
    borderline west={6pt}{0pt}{yellow},
    colback=yellow!10!white}
\newtcolorbox{importantBox}{textmarker,
    borderline west={6pt}{0pt}{red},
    colback=red!10!white}
\newtcolorbox{noteBox}{textmarker,
    borderline west={8pt}{0pt}{gray},
    colback=gray!10!white}
\theoremstyle{definition}
\newtheorem{definition}{Definition}[subsection]

\theoremstyle{remark}

\usetikzlibrary{arrows.meta}

\usepackage{tikz}
\usetikzlibrary{plotmarks}
\usetikzlibrary{arrows,shapes,positioning}
\usetikzlibrary{decorations.markings}
\tikzstyle arrowstyle=[scale=1]
\tikzset{>=latex}

\definecolor{chestnut}{rgb}{0.8, 0.36, 0.36}

\newcommand{\maintain}{cli\&lib-maintain}
\newcommand{\client}{cli-maintain}
\newcommand{\library}{lib-maintain}
\newcommand{\nonmaintain}{non-maintain}
\newcommand\ecocon{DC congruence}

\AtBeginDocument{%
  \providecommand\BibTeX{{%
    \normalfont B\kern-0.5em{\scshape i\kern-0.25em b}\kern-0.8em\TeX}}}

\begin{document}
\title{Giving Back: Contributions Congruent to Library Dependency Changes in a Software Ecosystem}

\author{Supatsara~Wattanakriengkrai, Dong~Wang, Raula~Gaikovina~Kula, Christoph~Treude,  Patanamon~Thongtanunam, Takashi~Ishio, and Kenichi Matsumoto
\IEEEcompsocitemizethanks{
\IEEEcompsocthanksitem S. Wattanakriengkrai, R. Kula, T. Ishio, and K. Matsumoto are with Nara Institute of Science and Technology, Japan.\newline  E-mail: \{wattanakri.supatsara.ws3, raula, ishio, matumoto\}@is.naist.jp.
\IEEEcompsocthanksitem D. Wang is with Kyushu University, Japan.\newline  
E-mail: wang.dong.vt8@is.naist.jp.
\IEEEcompsocthanksitem C. Treude and P. Thongtanunam are with the University of Melbourne, Australia.\newline  E-mail: \{christoph.treude, patanamon.t\}@unimelb.edu.au.
}
}

\IEEEtitleabstractindextext{%
\begin{abstract}
Popular adoption of third-party libraries for contemporary software development has led to the creation of large inter-dependency networks, where sustainability issues of a single library can have widespread network effects.
Maintainers of these libraries are often overworked, relying on the contributions of volunteers to sustain these libraries.
% In this work, we hypothesize that the developers who rely on a library because their code depends on it might also be the ones who contribute to the library, i.e., their contributions and dependencies are aligned in an ecosystem.
In this work, we measure contributions that are aligned with dependency changes, to understand where they come from (i.e., non-maintainer, client maintainer, library maintainer, and library and client maintainer), analyze whether they contribute to library dormancy (i.e., a lack of activity), and investigate the similarities between these contributions and developers' typical contributions.
Hence, we leverage socio-technical techniques to measure the dependency-contribution congruence (DC congruence), i.e., the degree to which contributions align with dependencies. 
We conduct a large-scale empirical study to measure the \ecocon~for the NPM ecosystem using 1.7 million issues, 970 thousand pull requests (PR), and over 5.3 million commits belonging to 107,242 NPM packages. 
At the ecosystem level, we pinpoint in time peaks of congruence with dependency changes (i.e., 16\% DC congruence score).
Surprisingly, these contributions came from the ecosystem itself (i.e., non-maintainers of either client and library).
At the project level, we find that \ecocon~shares a statistically significant relationship with the likelihood of a package becoming dormant.
Finally, by comparing source code of contributions, we find that congruent contributions are statistically different to typical contributions. 
Our work has implications to encourage and sustain contributions, especially to support library maintainers that require dependency changes.

\end{abstract}

}

\maketitle

\IEEEdisplaynontitleabstractindextext

\IEEEpeerreviewmaketitle

\input{sections/1intro.tex}
\input{sections/2background.tex}

\input{sections/3setup}

\input{sections/4results.tex}

\input{sections/5related_work.tex}

\ifCLASSOPTIONcaptionsoff
  \newpage
\fi

\bibliographystyle{IEEEtranS}
\bibliography{IEEEabrv,filteredref.bib}

\end{document}

%% file: sections/1intro.tex
\section{Introduction}
The adoption of third-party libraries for contemporary software development has led to the emergence of massive library platforms (i.e., package ecosystems) such as NPM for JavaScript\footnote{\url{https://www.npmjs.com/}} which is reported to be relied upon by more than 11 million developers worldwide, and contains more than one million packages \cite{npm}. Other examples include the 427,286 Maven libraries for Java Virtual Machine languages,\footnote{https://search.maven.org/stats} and the 324,779 PyPI packages for the Python community,\footnote{\url{https://pypi.org/}} to name a few. These package ecosystems comprise of a complex inter-connected network of dependencies, where developers adopt many other packages for their work. As an example, NPM packages directly depend on between 5 to 6 other packages in the ecosystem on average~\cite{kula2017impact, wittern2016look}.

The reliance on other packages in often brittle dependency chains~\cite{10.1109/MSR.2017.55} implies that local sustainability issues around individual packages can have widespread network effects~\cite{fse2018_sustained}. Contributors and maintainers of these packages are often overworked volunteers, who can decide to stop contributing at any time~\cite{robles2006contributor}. Sustained contributions are crucial to be able to respond to breaking changes from elsewhere in the ecosystem or apply critical security patches~\cite{zapata2018towards}. 
In response to recent attacks on open source libraries,\footnote{Article at \url{https://www.zdnet.com/article/google-here-comes-our-open-source-maintenance-crew/}} Google has deployed an "Open Source Maintenance Crew", tasked to assist upstream maintainers of critical open-source libraries that are used by major technology vendors, including Microsoft, Google, IBM and Amazon Web Services. 
Yet, the extent, origins and motivations behind these contributions are unclear.

To gain intuition about the motivations of contributors, we first contacted a few Node.js developers to ask \textit{why they contribute to other NPM packages, i.e., packages which they do not own?}
Most of them responded that they contributed because their code depends on those packages, e.g.,
\textit{``I rely on them [other packages] or want to rely on them and need to add/fix something in order to do so.''} and
\textit{``When they're broken and need fixing (and it's blocking my work), that's usually when I jump in.''}
One developer also noted that the ultimate goal of making contributions is to support the entire ecosystem:
\textit{``In my experience with the Open Source world, the line between making contributions to my own projects or to other's becomes diffuse. Everybody is looking for the same: making better systems.''}
These responses motivated us to hypothesize that contributions congruent with dependency changes at the project level, could be aligned and detected at ecosystem level.
In other words, we hypothesize that contributions and dependency changes are aligned in a package ecosystem.

Hence, in this work, we set out to investigate how developer contributions align
with package dependencies in an ecosystem. To measure this alignment, we adopt the concept of socio-technical congruence which measures the alignment between task dependencies among people and coordination activities. Borrowing this concept, we measure the dependency-contribution (DC) congruence, i.e., the degree to which the contributions align with dependencies.
Through the case of the NPM ecosystem, we conduct a large scale empirical study to examine the \ecocon~and the package dormancy using over 5.3 million commits, 1.7 million issues, and 970 thousands pull requests (PRs) belonging to 107,242 packages. We formulate the following research
questions.

\noindent
$\bullet$~{{\bf (RQ1)} \it How do developer contributions align with their dependencies? }
\textit{Motivation:}
The motivation for the first research question is to examine the extent of the alignment at the ecosystem level, and confirm our hypothesis. We focus on the following sub-questions: 
\noindent\textit{RQ1a.} \textit{To what degree do developer contributions align with dependencies?} and
\noindent\textit{RQ1b.} \textit{How do different contribution types and dependency changes contribute to \ecocon?}
\textit{Results:}
The alignment between developer contributions and dependency changes varies over time, with the highest values generally observed in earlier stages of the NPM ecosystem. The peaks we identify in the data correspond to a \ecocon~of 0.16, i.e., 16\% of the dependency changes receive aligned contributions.
These contributions are most aligned with downgrading a dependency, and mainly come from developers that are not maintainers of both the client and library.

\noindent
$\bullet$~{{\bf (RQ2)} \it What is the relationship between the \ecocon~and the likelihood of packages becoming dormant?}
\textit{Motivation:}
Since package sustainability heavily relies on contributions \cite{Samoladas:IST2010}, we would like to investigate the association between the \ecocon~and the package dormancy.
  
\textit{Results:}
Our survival analysis shows that the various types of \ecocon~share an inverse relationship with the likelihood that a package becomes dormant, i.e., the lower the \ecocon, the more likely the package becomes dormant.
For instance, packages that received more congruent contributions (i.e., issues) from non-maintainers are less likely to become dormant.
\\
\noindent
$\bullet$~{{\bf (RQ3)} \it Do the contributions differ depending on their alignment with dependencies?} 
\textit{Motivation:}
In addition to measuring \ecocon~(RQ1 and RQ2), since developers may be motivated to submit aligned contributions based on their dependencies, for RQ3, we would like to confirm the extent to which these contributions differ from their typical contributions. Hence, we conduct a deeper investigation on the similarity of contributions given different types of contributions identified in the first two research questions.

\textit{Results:}
Comparing contribution similarity in terms of source code and file paths, we find statistical differences in source code content of aligned contributions submitted to dependencies when compared to those that are not aligned.
In other words, these congruent contributions are not the typical source code submitted by that contributor.

We provide an online appendix, containing datasets and source code related to (a) the ecosystem-level and package-level \ecocon~results, (b) the metric data for our survival model analysis, and (c) the file path and source code similarities between contribution types, which is available at \url{https://doi.org/10.5281/zenodo.5677371}.

%% file: sections/2background.tex
\section{Dependencies and Contributions}
\label{sec:background}
In this section, we describe how we measure the \ecocon, i.e., the degree to which contributions align with dependencies as a graph.
The graph is temporal, i.e., it is constructed based on a time period.

\begin{table*}[t]
    
\centering

\caption{Classification of contributions by maintainer role to both client and library packages.}
\scalebox{0.9}{
\label{tab:edgeattributes}
\begin{tabular}{@{}lll@{}}

\toprule
\textbf{Contribution Type} & \textbf{Description} & \textbf{Algorithm} \\ \midrule
\maintain & Contributions from a developer who commits to both client and library packages. & \begin{tabular}[c]{@{}l@{}}\footnotesize{$E_{c-c} = \{ N_{pkg.c} \rightarrow  N_{pkg.l} \in E_{t1-t2} ~ |$}\\ \footnotesize{$\exists N_{dev.d}.~ N_{dev.d} \xRightarrow{c} N_{pkg.c} \land N_{dev.d} \xRightarrow{c} N_{pkg.l} \}$}\end{tabular} \\ 
\client & Contributions from a developer who commits to a client and submits to a library. & \begin{tabular}[c]{@{}l@{}}\footnotesize{$E_{c-s} = \{ N_{pkg.c} \rightarrow  N_{pkg.l} \in E_{t1-t2} ~ |$}\\ \footnotesize{$\exists N_{dev.d}.~ N_{dev.d} \xRightarrow{c} N_{pkg.c} \land N_{dev.d} \xRightarrow{s} N_{pkg.l} \}$}\end{tabular}  \\
\library & Contributions from a developer who submits to a client and commits to a library. & \begin{tabular}[c]{@{}l@{}}\footnotesize{$E_{s-c} = \{ N_{pkg.c} \rightarrow  N_{pkg.l} \in E_{t1-t2} ~ |$}\\ \footnotesize{$\exists N_{dev.d}.~ N_{dev.d} \xRightarrow{s} N_{pkg.c} \land N_{dev.d} \xRightarrow{c} N_{pkg.l} \}$}\end{tabular} \\
\nonmaintain & Contributions from a developer who submits to both client and library packages. & \begin{tabular}[c]{@{}l@{}}\footnotesize{$E_{s-c} = \{ N_{pkg.c} \rightarrow  N_{pkg.l} \in E_{t1-t2} ~ |$}\\ \footnotesize{$\exists N_{dev.d}.~ N_{dev.d} \xRightarrow{s} N_{pkg.c} \land N_{dev.d} \xRightarrow{s} N_{pkg.l} \}$}\end{tabular} \\ \bottomrule
\end{tabular}
}
\end{table*}

\begin{figure}[t]
\begin{subfigure}{\linewidth}
    \centering
  
    \begin{tikzpicture}[
        roundnode/.style={circle, fill=black, minimum size=5mm},
        squarenode/.style={fill=black, text=red, minimum size=5mm},
    ]
    \begin{scope}
        
        \node[roundnode, label=above:$N_{pkg.i}$] (s2_proji) at (3, 2.5) {};
        \node[roundnode, label=below:$N_{pkg.j}$] (s2_projj) at (4,0) {};
        \node[roundnode, label=below:$N_{pkg.k}$] (s2_projk) at (2,0) {};
      
        \node[squarenode, label=below:$N_{dev.x}$] (s3_devx) at (6, 1.25) {};

    \end{scope}

    \begin{scope} [every edge/.style={draw=gray, very thick}]
        \path [->] (s2_proji) edge (s2_projj);
        \path [->] (s2_proji) edge (s2_projk);
        \path [->] (s2_projj) edge (s2_projk);
     
    \end{scope}
    \begin{scope} [every edge/.style={draw=gray, thick, double distance=2pt}]
       
        \path [->] (5.7, 1.25) edge node[right = 3mm] {$c$} (s2_proji);
        \path [->] (5.65 , 1.25) edge node[below = 1mm] {$c$} (s2_projj);
    
    \end{scope}

    \end{tikzpicture}
    
    \caption{Nodes and edges from time period Oct 1 to Dec 31, 2015}
    \label{fig:graph_before}
    \vspace{2ex}
\end{subfigure}
\begin{subfigure}{\linewidth}
    \centering

    \begin{tikzpicture}[
        roundnode/.style={circle, fill=black, minimum size=5mm},
        squarenode/.style={fill=black, text=red, minimum size=5mm},
    ]
    \begin{scope}
        
        \node[roundnode, label=right:$N_{pkg.i}$] (s2_proji) at (4, 1.5) {};
        \node[roundnode, label=below:$N_{pkg.j}$] (s2_projj) at (5,0) {};
        \node[roundnode, label=below:$N_{pkg.k}$] (s2_projk) at (3,0) {};
        
        \node[squarenode, label=below:$N_{dev.x}$] (s3_devx) at (2, 3.5) {};
        
        \node[squarenode, label=below:$N_{dev.y}$] (s3_devy) at (6, 3.5) {};
        
    \end{scope}

    \begin{scope} [every edge/.style={draw=gray, very thick}]
        \path [->] (s2_proji)  edge  (s2_projj);
        \path [->] (s2_proji) edge (s2_projk);
        \path [->] (s2_projj) edge (s2_projk);
        
    \end{scope}
    \begin{scope} [every edge/.style={draw=gray, thick, double distance=2pt}]
        \path [->] (6, 2.8) edge node[left = 2mm] {$c$} (s2_proji);
        \path [->] (6, 2.8) edge[bend left=15] node[right = 1mm] {$s$} (s2_projj);
        \path [->] (2, 2.8) edge node[right = 2mm] {$c$} (s2_proji);
        \path [->] (2, 2.8) edge[bend right=15] node[left = 1mm] {$c$} (s2_projk);
    \end{scope}

    \end{tikzpicture}
    
    \caption{Node and Edges from time period Jan 1 to Mar 31, 2016}
    \label{fig:graph_in_period}
\end{subfigure}
\begin{subfigure}{\linewidth}
    \centering

    \begin{tikzpicture}[
        roundnode/.style={circle, fill=black, minimum size=5mm},
        squarenode/.style={fill=black, text=red, minimum size=5mm},
    ]
    \begin{scope}
        
        \node[roundnode, label=right:$N_{pkg.i}$] (s2_proji) at (4, 1.5) {};
        \node[roundnode, label=below:$N_{pkg.j}$] (s2_projj) at (5,0) {};
        \node[roundnode, label=below:$N_{pkg.k}$] (s2_projk) at (3,0) {};
        
    \end{scope}

    \begin{scope} [every edge/.style={draw=gray, very thick}]
        \path [->][dashed] (s2_proji)  edge node[right = 2mm] {cli-maintain} (s2_projj);
        \path [->][dashed] (s2_proji) edge node[left= 2mm] {cli\&lib-maintain} (s2_projk);
    
    \end{scope}

    \end{tikzpicture}
    
    \caption{Removing the developer nodes, we transform Fig.~1b to create the contribution relationships between packages from time period Jan 1 to Mar 31, 2016}
    \label{fig:graph_for_congruence}
\end{subfigure}
\caption{Example graph representation of the relationships between contributions and dependencies}
 \label{fig:graph}
\end{figure}
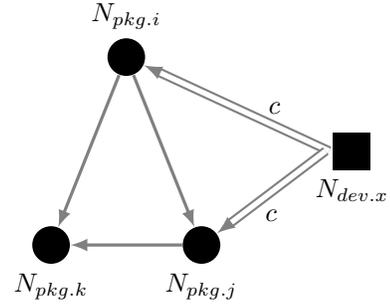
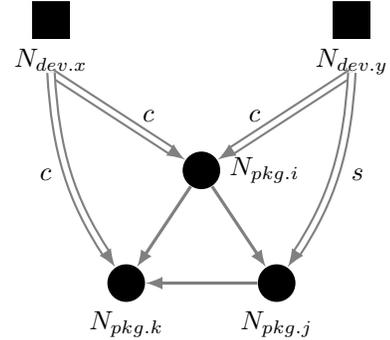
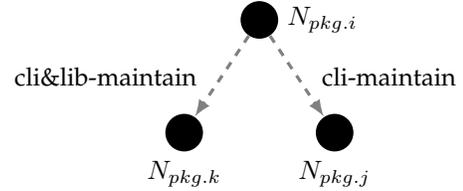

\subsection{A Dependency-Contribution Graph}
Figure \ref{fig:graph} shows a visual example to illustrate a Dependency-Contribution graph (DC graph) that captures the relationship between contributions and dependencies in the ecosystem.
Let $G = (N, E)$ be a directed graph where $N$ represents a set of nodes and $E$ represents a set of edges.
$N$ has two types of nodes (i.e., $N_{dev}$ and $N_{pkg}$), where $N_{dev}$ refers to a developer, while $N_{pkg}$ refers to a software package.
$N_{dev}$ has an attribute $<$userid$>$ where userid is the unique identifier of the developer.
$N_{pkg}$ has the attributes $<$name, time$>$ where name is the unique name of the package and time is when the package was first created.

The edges ($E$) represent the relationships between nodes in terms of contributions and dependencies. Hence, we define two edge types in our graph, i.e., package dependencies and contributions to packages (see details in Sections
2.2 and 2.3). 
The graph edges have a temporal attribute which provides information when the dependency (or contribution) event occurs. 
Since the contribution and dependency relationships are dynamic (i.e., changing over time), we capture and analyze these relationships based on a specific period of time.
Hence, we defined a temporal graph $G_{t1-t2}$ where t1 and t2 are between two time points which we capture the contribution and dependency relationships.
The temporal graph $G_{t1-t2}$ is formally defined as:
   $ G_{t1-t2} = (N_{t1-t2}, E_{t1-t2})$
where $N_{t1-t2}$ is a set of package and developer nodes that have relationships with others within a period of $t1$ until $t2$, and $E_{t1-t2}$ is a set of edges where the contributions or dependency events (e.g., a dependency change) between the packages in $N_{t1-t2}$ occur during the period of $t1$ until $t2$.
It is important to note that for any time-period (i.e., t1 to t2), a package node must be created before t1.
This means that any package created in one time period will only be considered in the next time period.

Figure \ref{fig:graph} provides illustrative examples of two subsequent time periods (i.e., $G_{Oct2015-Dec2015}$ and $G_{Jan2016-Mar2016}$).
Specifically, $G_{Oct2015-Dec2015}$ captures the relationships between nodes in the time period of 1 October until 31 December 2015 which shows that only a developer $N_{dev.x}$ contributed to package $N_{pkg.i}$ and $N_{pkg.j}$, note that $N_{pkg.i}$ depends on $N_{pkg.j}$ and $N_{pkg.k}$, and $N_{pkg.j}$ depends on $N_{pkg.k}$ as shown by the edges in the graph. Then in, $G_{Jan2016-Mar2016}$, developer $N_{dev.x}$ contributed to packages $N_{pkg.i}$ and $N_{pkg.k}$, and another developer $N_{dev.y}$ contributed to $N_{pkg.i}$ and $N_{pkg.j}$.

\subsection{Graph Edges: Package Dependencies}
\begin{definition}[Edge:  $N_{pkg} \rightarrow N_{pkg}$]
The first edge type represents dependency relationships between two packages.
For example, $N_{pkg.i} \rightarrow N_{pkg.j}$ means that a package $N_{pkg.i}$ uses (i.e., depends on) a package  $N_{pkg.j}$, where $N_{pkg.i}$ is a client to the $N_{pkg.j}$ library.
\end{definition}

\textbf{Edge Attributes: Dependency changes}.
The dependency edge has the attributes $<$change, time$>$ which is the current change of the dependency relationship.
We capture the edge dependency change within a time period (between $t1-t2$). 
It is important to note that the edge of dependency change that occurs in a time period will be kept in the subsequent time periods until a new dependency event occurs.
In other words, the latest dependency change is carried over into the next time period.
We categorize the dependency change into four types
: (i) is-added, i.e., a dependency has been added to a package, (ii) is-removed, i.e.,  the dependency has been removed from a package, (iii) is-upgraded, i.e., a dependency version has a been changed to a more recent version of a library, and (iv) is-downgraded, i.e., a dependency version has been changed to an older version of a library.

\subsection{Graph Edges: Contributions to Package}
\begin{definition}[Edge: $N_{dev}  \xRightarrow{} N_{pkg}$]
The second edge type represents the contribution relationship between a developer and a package.
For example, $N_{dev.x}  \xRightarrow{} N_{pkg.i}$ means that a developer $N_{dev.x}$ contributes to a package $N_{pkg.i}$. 
\end{definition}

\textbf{Edge Attributes: Developer roles}.
The contribution edge has two attributes $<$role, time$>$ where the time is when the contribution was made.
The role refers to the access levels of the developer to the package when making a contribution. 
There are two developer roles: a \texttt{committer($c$)}, who has the ability to merge any submitted contributions, and a \texttt{submitter($s$)}, who can only submit contributions to a package but does not have the permission to merge the contribution.

We now define four types of contributions between packages.
When capturing these  contribution relationships between packages, as a minimum, we capture developers who made at least two contributions to the two different packages that have a dependency relationship between them.
Formally, given a dependency relationship between the client
and library packages, 
we define the types of contribution (see Table \ref{tab:edgeattributes}) as follow: (i) \maintain, i.e., contributions from a developer who is a committer to both client and library packages, (ii) \client, i.e., contributions from a developer who is a committer to a client that submits to a library, (iii) \library, i.e., contributions from a developer who is a submitter to a client that commits to a library, (iv) \nonmaintain, i.e., contributions from a developer who is a submitter to both client and library packages.
Therefore, the DC graph in Figure \ref{fig:graph}b can be transformed to the contribution graph in Figure \ref{fig:graph}c, where $N_{dev.y}$ commits to $N_{pkg.i}$ and $N_{dev.y}$ submits to $N_{pkg.j}$, making these contributions a \client~type.

\subsection{Congruence Calculation}
To measure the \ecocon, 
we use the adjacency matrix to capture the contribution and dependency relationships in a temporal graph $G_{t1-t2}$.

\textbf{Matrix Construction}.
Based on the contribution and dependency edges, we calculate the congruence at two levels: 1) ecosystem level and 2) package level.
Specifically, we construct a contribution matrix ($C_E$) and a dependency matrix ($C_D$). 
We then measure the congruence between the contributions and dependencies based on the definition of socio-technical congruence by Cataldo et al.~\cite{Cataldo:2008ESEM} and Wagstrom et al.~\cite{doi:10.5465/ambpp.2010.54500789}.
Briefly stated, they defined socio-technical congruence as the
match between the coordination requirements established by the dependencies among tasks and the actual coordination activities performed by the developers.
We provide a formal definition for each matrix and congruence calculation as shown below.

\begin{definition}[$C_D$ adjacency matrix]
The dependency matrix ($C_D$) of a $G_{t1-t2}$ captures the relationships between packages in $N_{t1-t2}$ based on the dependency changes in $E_{t1-t2}$.
The $C_D$ matrix has a size of $|N_{t1-t2}| \times |N_{t1-t2}|$ where its elements indicate whether there is a dependency relationship between the two packages or not.
In other words, $C_D[i,j] =  1$ if there exists an edge $N_{pkg.i} \rightarrow N_{pkg.j} \in E_{t1-t2}$, otherwise 0.
For example, the graph $G_{Jan2016-Mar2016}$ in Figure \ref{fig:graph_in_period} has three dependency relationships, i.e., $N_{pkg.i} \rightarrow N_{pkg.j}$, $N_{pkg.i} \rightarrow N_{pkg.k}$, and $N_{pkg.j} \rightarrow N_{pkg.k}$.
Hence, the dependency matrix $C_D$ of $G_{Jan2016-Mar2016}$  can be constructed as:
\[
 \footnotesize{
\begin{blockarray}{cccccc}
 & N_{pkg.i} & N_{pkg.j} & N_{pkg.k} \\
\begin{block}{c[ccccc]}
N_{pkg.i} & 0 & 1& 1 \bigstrut[t] \\
N_{pkg.j} & 0 &0 & 1  \\
N_{pkg.k} & 0 &0 & 0  \\
\end{block}
\end{blockarray}\vspace*{-1.25\baselineskip}
}
\]
\end{definition}
Furthermore, we calculate the $C_D$ adjacency matrix for each type of dependency change (as described in Section 2.2), resulting in a total of four matrices.

\begin{definition}[$C_E$ adjacency matrix]
The contribution matrix ($C_E$) of a $G_{t1-t2}$ captures the relationship between packages in $N_{t1-t2}$ based on the contributions of the associated developers in $E_{t1-t2}$.
The $C_E$ matrix has a size of $|N_{t1-t2}|$ x $|N_{t1-t2}|$ where its elements indicate whether there are contributions between two packages or not.
In other words, $C_E[i,j] = 1$ if there exists an edge between any two packages for any developer contribution, otherwise 0.
Note that to create the contribution matrix, we use the contribution graph (like in Figure \ref{fig:graph}c) which is
transformed from the DC graph (like Figure \ref{fig:graph}b) based on the developer roles and dependency
relationships.
For example, the contribution matrix $C_E$ for the \client~contribution of the graph $G_{Jan2016-Mar2016}$ in Figure \ref{fig:graph}c can be constructed as:
\[
 \footnotesize{
\begin{blockarray}{cccccc}
 &N_{pkg.i} & N_{pkg.j} & N_{pkg.k} \\
\begin{block}{c[ccccc]}
N_{pkg.i} & 0 & 1& 0 \bigstrut[t] \\
N_{pkg.j} & 0 &0 & 0 \\
N_{pkg.k} & 0 &0 & 0 \bigstrut[b]\\
\end{block}
\end{blockarray}\vspace*{-1.25\baselineskip}
}
\]
\end{definition}
where there is a committer ($N_{dev.y}$) of a client package ($N_{pkg.i}$) contributing to a library package ($N_{pkg.j}$).
We calculate the $C_E$ adjacency matrix for each type of contribution (as described in Section 2.3), resulting in a total of four matrices.

\begin{definition}[Ecosystem-level DC Congruence] 
To measure the alignment between contributions and dependencies of packages at the ecosystem level, we adapt the definition of socio-technical congruence by Cataldo \cite{Cataldo:2008ESEM} as follows:

\begin{equation}
\begin{multlined}
   EcosystemC(G_{t1-t2}) =  \frac{\Sigma (C_D \wedge C_E)}{\Sigma C_D}
\end{multlined}
\end{equation}
In sum, Ecosystem-level DC Congruence is the ratio of dependencies that receive aligned contributions divided by the total number of dependencies in the ecosystem.
For example, given $G_{Jan2016-Mar2016}$ in Figure \ref{fig:graph_in_period} and all the dependency relationships in the graph have a type of ``is.added'', we want to calculate the \ecocon~between dependencies of is.added and \client~contributions. 
Hence, $\Sigma (C_D \wedge C_E) = 1$ as there is a \client~contribution relationship that matches with the dependency relationship (i.e., $N_{pkg.i} \rightarrow  N_{pkg.j}$) between the DC graph (in Figure \ref{fig:graph_in_period}) and the contribution graph (in Figure \ref{fig:graph_for_congruence}).
Finally, the \ecocon~at the ecosystem level of the graph $G_{Jan2016-Mar2016}$ is $EcosystemC(G_{Jan2016-Mar2016})  = \frac{1}{3} = 0.333$.

\textbf{DC Congruence Values}.
For each time period, we construct the matrices ($C_D$ and $C_E$) for each dependency change (cf. Table \ref{tab:edgeattributes}) and the four contribution types (i.e., \maintain, \client, \library, and \maintain~contributions).
Hence, for each time period, we calculate a total of 16 values of \ecocon~(four $C_D$ matrices multiplied by four $C_E$ matrices).
\end{definition}

\begin{definition}[Package-level DC Congruence]
To quantify how the congruence related to a package contributed to the congruence value of the whole ecosystem, we also measure congruence of each package (i.e., the congruence at the package level).
To do so, we adapt the individualized socio-technical congruence definition of Wagstrom \cite{doi:10.5465/ambpp.2010.54500789} as follows:

\begin{equation}
\begin{multlined}
   PgkC(G_{t1-t2}, p) = \frac{ \Sigma (C_{D}[p,] \wedge C_{E}[p,])  + \Sigma (C_{D}[,p] \wedge C_{E}[,p])}{\Sigma (C_{D}[p,]) + \Sigma (C_{D}[,p])}
\end{multlined}
\end{equation}
where $p$ is a package of interest (N$_{pkg.p}$) in the graph $G_{t1-t2}$.  
The numerator calculates the number of dependencies for package $p$ that receive aligned contributions over both the rows and columns for that package. The denominator
simply sums up the number of dependencies incident upon package $p$.
For example, given $G_{Jan2016-Mar2016}$ in Figure \ref{fig:graph_in_period} and all the dependency relationships in the graph have a type of ``is.added'', we want to calculate the package-level \ecocon~of the package $N_{pkg.k}$ between dependencies of is.added and \maintain~contributions. 
Hence, $\Sigma (C_{D}[N_{pkg.k},]) + \Sigma (C_{D}[,N_{pkg.k}])$ = 2 as the package $N_{pkg.k}$ has two dependency relationships (i.e., $N_{pkg.i} \rightarrow  N_{pkg.k}$ and $N_{pkg.j} \rightarrow  N_{pkg.k}$), and
$\Sigma (C_{D}[N_{pkg.k},] \wedge C_{E}[N_{pkg.k},])  + \Sigma (C_{D}[,N_{pkg.k}] \wedge C_{E}[,N_{pkg.k}])$ = 1 as there is a \maintain~contribution relationship that
matches with the dependency relationship (i.e., $N_{pkg.i} \rightarrow  N_{pkg.k}$) between the DC graph (in Figure \ref{fig:graph_in_period}) and the contribution graph (in Figure \ref{fig:graph_for_congruence}).
Finally, the package-level congruence of the package $N_{pkg.k}$ is calculated as follows.
\begin{align*}
PgkC( G_{Jan2016-Mar2016}, N_{pkg.k})  = \frac{1}{2} = 0.5
\end{align*}
\end{definition}

Intuitively, we can interpret the congruence values (0 to 1) as the extent to which contributions align with dependency changes. 
A high congruence value at the ecosystem level would indicate that a large number of dependency relationships receive aligned contributions, while a low value would indicate that there is a small number of aligned contributions in the ecosystem.
Similarly, at the individual level, a higher congruence value would indicate that a large number of contributions aligned with dependency relationships of that package.

%% file: sections/3setup.tex
\begin{table}[t]
\caption{Dataset snapshot statistics.}
\label{tab:data-statistic}
\begin{tabular}{lc|rrrr}
\hline
\textbf{}                     & \textbf{\begin{tabular}[c]{@{}c@{}}\# N$_{pkg}$ \end{tabular}} & \multicolumn{1}{l}{\textbf{\# PRs}}  & \multicolumn{1}{l}{\textbf{\# issues}} & \textbf{\# commits} & \multicolumn{1}{l}{\textbf{\# dev}} \\ \hline
2014                          & 17,859                                                              & 22,239                               & 55,081                                 & 473,591             & 11,031                                     \\
2015                          & 57,534                                                              & 98,975                               & 222,146                                & 1,255,065           & 46,029                                     \\
2016                          & 92,748                                                              & 242,424                              & 427,001                                & 1,415,015           & 82,991                                     \\
2017                          & 105,058                                                             & 186,257                              & 352,411                                & 886,796             & 96,916                                     \\
2018                          & 106,815                                                             & 139,016                              & 267,318                                & 527,252             & 86,978                                     \\
2019                          & 107,146                                                             & 166,588                              & 259,560                                & 428,191             & 69,285                                     \\
2020                          & 107,242                                                             & 115,186                              & 171,435                                & 339,219             & 43,815                                     \\ \hline
\multicolumn{1}{c}{\textbf{}} & \textbf{Total}                                                      & \multicolumn{1}{l}{\textbf{970,685}} & \multicolumn{1}{l}{\textbf{1,754,952}} & \textbf{5,325,129}  & \textbf{437,045}                           \\ \hline
\end{tabular}
\end{table}

\section{Data Preparation}
\label{sec:data_preparation}

To evaluate our congruence models and answer our RQs, we perform an empirical study on the NPM package ecosystem.
We selected the NPM JavaScript ecosystem as it is one of the largest package collections on GitHub, and also has been the focus of recent studies \cite{10.1109/MSR.2017.55, 10.1145/3106237.3106267, decan2017empirical}.
In this work, we focus on the NPM packages linked to their GitHub repositories, since we will analyze developer contributions (i.e., PRs and issues).
Similar to `all-the-package-repos' \cite{GitHubni50:online} and Chinthanet et al.~\cite{chinthanet2021lags}, we use a listing of NPM packages from the NPM registry and then matched them to repositories available on GitHub.
The listing of NPM packages was acquired in Nov. 2020.
To collect the required information, we use the GitHub Rest API\footnote{https://docs.github.com/en/rest} to collect PRs, issues, commits, and other metadata (e.g., license type) of the package repositories included in the latest data snapshot.
The issues are used to answer RQ1 and RQ2, while the commits are used to answer RQ2 and the PRs are used to answer all RQs.
To overcome the GitHub API limit, we use five GitHub tokens that were generated by the first five authors during the collection process.
With these tokens, the API rate limit is 25,000 (5,000 * 5) requests per hour.
We took one month, from November 1 to November 30, 2020, to acquire all information.

Table \ref{tab:data-statistic} provides an overview of the number of NPM packages that have repositories available on GitHub for each year.
Overall, the number of NPM packages starts from small (i.e., 17,859 packages), but then reaches over a hundred thousand packages (i.e., 107,242) as of Nov. 2020. 
Similar to previous work which considered the temporal dimension \cite{Mirsaeedi:icse2020, Brindescu:emse2020, Nassif:icsme2017}, we split the data of each year into quarters:  (Q1) January 1 to March 31, (Q2) April 1 to June 30,  (Q3) July 1 to September 30, and (Q4) October 1 to December 31 to establish the temporal intervals of our graph (i.e., $G_{t1-t2}$).

\textbf{Extracting Contributions, Roles and Dependencies}
Once the time-intervals were established, we then extract the developer contributions and dependency changes within each quarter.
For the contribution activities, we generate the graph nodes by identifying PRs and issues that occurred during the quarter.
To classify the role of the contributor, we use the GitHub Rest API to identify whether or not the developer who submitted a contribution (i.e., a PR or an issue) had committed any prior commits into the main branch of the repository. 
For the dependencies and their changes, similar to prior work \cite{Cogo:TSE2019}, we parse the \texttt{package.json} file to extract both development and runtime dependencies for each package during that time-period.
To identify the dependency changes during the time-period, we use the PyDriller package\footnote{https://github.com/ishepard/pydriller} to mine the dependency changes (cf. Table \ref{tab:edgeattributes}) of \texttt{package.json} in the Git history.
More specifically, we focus on the Modification object provided by the PyDriller to investigate commits to \texttt{package.json}.

\textbf{Calculating Ecosystem-Level DC Congruence (RQ1)}
Answering RQ1 involves computing the \ecocon~at the ecosystem level.
Hence, for all time-periods, we construct $C_E$ and $C_D$ matrices of all the combinations of contribution types against the different dependency changes for PRs and issues separately (see Section 2.4).
To ease computational power, we filter out all rows and columns of the $C_E$ and $C_D$ matrices that have only zeros. 
Since we analyze PRs and issues separately, we generated a total of 32 instances for each quarter, resulting in 864 matrices.
We took around 21 days of execution based on two  standard computing machines with an Intel Core 2.60GHz, with 40 cores and 252GB of RAM.

\textbf{Calculating Package-Level DC Congruence (RQ2)}
Answering RQ2 involves computing the package-level DC congruence.
Using the constructed matrices (before filtering the rows and columns with zeros) from RQ1, we calculate the package-level DC congruences for each time-period.
Similar to RQ1, we generated a total of 32 \ecocon~values for each package in each quarter.
%It is important to note that we automatically set the package-level DC congruence to zero if a package does not occur in the DC graph constructed during the time-period.
Since the experiment was conducted in parallel with RQ1, the computational time and resources are the same.

\textbf{Sampling Contributions (RQ3)}
Answering RQ3 involves a quantitative analysis of developer contributions. 
Hence, we drew a statistically representative random sample of 383 developers.
Our sample size allows us to generalize conclusions about the ratio of developers to all developers with a confidence level of 95\% and a confidence interval of 5\%, as suggested by prior work \cite{Hata:ICSE2019} based on the Sample Size Calculator.\footnote{https://www.surveysystem.com/sscalc.htm}
To collect the most recent contributions of the sampled developers, we extracted contributions belonging to the most recent quarter.
To do so, we first identified the last contribution, i.e., the last submitted PRs in our dataset, and then we extracted all their PRs submitted within that time-period.
In the end, we collected 8,938 PRs.
Since our approach involves similarity analysis of the source code (at least one .js file) and file-path components, we obtained 3,714 (42\%) PRs from the 383 developers with the four developer contribution types (see Section 2.3).

%% file: sections/4results.tex
\section{Results}
In this section, we present the results to answer our RQs.

\begin{table}[t]
\centering
\caption{Summary statistics of ecosystem-level \ecocon~of contribution types.}
\label{tab:rq1a_result}
\begin{tabular}{@{}cllll@{}}
\toprule
\textbf{}        \textbf{Contribution Type}     & \textbf{Max} & \textbf{Min} & \textbf{Mean} & \textbf{Median} \\ \midrule
\nonmaintain & 0.166      & 0.001       & 0.030        & 0.020          \\
\maintain& 0.095       & 0.000       & 0.013        & 0.006          \\
\client     & 0.166      & 0.000       & 0.006        & 0.003          \\
\library    & 0.083      & 0.000       & 0.003        & 0.002          \\ \bottomrule
\end{tabular}
\end{table}

\begin{figure*}[t]
\centering
\includegraphics[width=\linewidth]{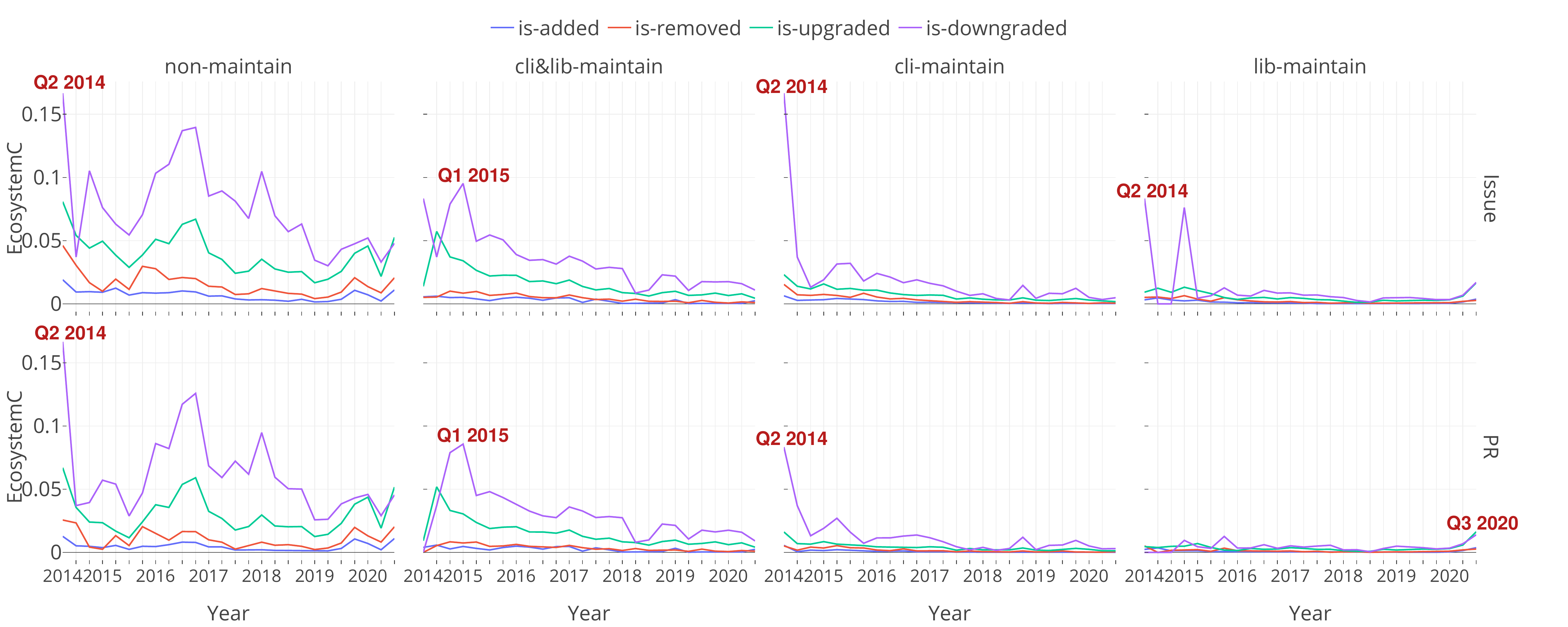}
\caption{Ecosystem-level \ecocon~over time. The red text indicates quarters in which congruence values are highest.}
\label{fig:rq1_result}
\end{figure*}

\begin{table*}[t]
\centering
\caption{Correlations between different dependency changes for each type of contribution}
\label{tab:rq1-corr}
\scalebox{0.8}{
\begin{tabular}{@{}lcccccccc@{}}
\toprule
\textbf{} &
  \multicolumn{2}{c}{\textbf{\nonmaintain}} &
  \multicolumn{2}{c}{\textbf{\maintain}} &
  \multicolumn{2}{c}{\textbf{\client}} &
  \multicolumn{2}{c}{\textbf{\library}} \\ \midrule
\textbf{Dependency change to Dependency change} &
  \multicolumn{1}{c}{\textbf{\begin{tabular}[c]{@{}c@{}}PR \\ corr. strength\end{tabular}}} &
  \multicolumn{1}{c}{\textbf{\begin{tabular}[c]{@{}c@{}}issue\\ corr. strength\end{tabular}}} &
  \multicolumn{1}{c}{\textbf{\begin{tabular}[c]{@{}c@{}}PR\\ corr. strength\end{tabular}}} &
  \multicolumn{1}{c}{\textbf{\begin{tabular}[c]{@{}c@{}}issue\\ corr. strength\end{tabular}}} &
  \multicolumn{1}{c}{\textbf{\begin{tabular}[c]{@{}c@{}}PR\\ corr. strength\end{tabular}}} &
  \multicolumn{1}{c}{\textbf{\begin{tabular}[c]{@{}c@{}}issue\\ corr. strength\end{tabular}}} &
  \multicolumn{1}{c}{\textbf{\begin{tabular}[c]{@{}c@{}}PR\\ corr. strength\end{tabular}}} &
  \multicolumn{1}{c}{\textbf{\begin{tabular}[c]{@{}c@{}}issue\\ corr. strength\end{tabular}}} \\ \midrule
is-added to is-removed &
  Strong &
  Strong &
  Moderate &
  Strong &
  Strong &
  Very strong &
  Moderate &
  Strong \\
is-added to is-upgraded &
  Strong &
  Strong &
  Moderate &
  Strong &
  Very strong &
  Very Strong &
  Strong &
  Very strong \\
is-added to is-downgraded &
  Moderate &
  Moderate &
  Moderate &
  Strong &
  Strong &
  Strong &
  Weak &
  Weak \\
is-removed to is-upgraded &
  Strong &
  Strong &
  Strong &
  Strong &
  Strong &
  Very strong &
  Moderate &
  Strong \\
is-removed to is-downgraded &
  Moderate &
  Moderate &
  Strong &
  Strong &
  Moderate &
  Strong &
  Weak &
  Moderate \\
is-upgraded to is-downgraded &
  Strong &
  Strong &
  Strong &
  Moderate &
  Very Strong &
  Strong &
  Moderate &
  Moderate \\ \bottomrule
   \multicolumn{6}{l}{Correlation: Negligible is 0 to 0.09, Weak is  0.1 to 0.39, Moderate is 0.4 to 0.69, Strong is 0.7 to 0.89, Very strong is 0.9 to 1. }\\
\end{tabular}
}
\end{table*}

\subsection{Assistance from the Ecosystem (RQ1)}

\textit{Approach.} 
To answer RQ1, we analyze the \ecocon~in terms of developer contributions and dependency changes.
In particular, we address each of the sub-questions as follow:
To answer RQ1a: \textit{To what degree do developer contributions align with dependencies?}, we summarize the \ecocon~in terms of the distribution by contribution type.
To answer RQ1b: \textit{How do different contribution types and dependency changes contribute to \ecocon}, we analyze trends of \ecocon~related to different contribution types and dependency changes. 

For RQ1b, we use the Spearman rank-order correlation coefficient to analyze the correlations between the different dependency changes for each kind of contribution.
A correlation between 0 -- 0.09 is considered as negligible, 0.1 -- 0.39 as weak, 0.4 -- 0.69 as moderate, 0.7 -- 0.89 as strong, and 0.9 -- 1 as very strong. To analyze the Spearman rank-order correlation coefficients, we use the dataframe.corr() function of the \texttt{pandas} package.\footnote{\url{https://pandas.pydata.org/}}

\textit{\textbf{Aligning contributions to their dependencies (RQ1a).}}
Table \ref{tab:rq1a_result} shows that the congruence between contributions and dependency changes is low (i.e., 0.001 to 0.166) when compared to other social-technical applications, such as aligning coordination and organization architectures \cite{Syeed:OSS2013}.
However, relatively, we find that \nonmaintain~contributions that aligned with downgrading dependencies and \client~contributions that aligned with downgrading dependencies had the highest congruence values (\ecocon~of 0.16).

From  Figure \ref{fig:rq1_result}, we find that the highest congruence values occur generally in the second quarter of 2014, followed by the first quarter of 2015.
This could be explained by the ecosystem size (17,859 and 57,534 packages in 2014 and 2015 respectively).

\textit{\textbf{Attributes that correlate with the \ecocon~(RQ1b)}}
From Figure \ref{fig:rq1_result}, we are able to make two observations.
First, we find that there is higher congruence with contributions when a dependency is downgraded (see the purple lines in Figure \ref{fig:rq1_result}).
This finding is consistent between all contribution types and between issues and PRs, except for the congruence between \library~PRs and dependency upgrades.
The second observation is that the \nonmaintain~contributions are most congruent with dependency changes.
This is consistent from both the issue and PRs.
This provides evidence that package dependency changes are satisfied by contributions that come from the ecosystem and not the maintainers of either that package or clients.

Table \ref{tab:rq1-corr} shows evidence that all dependency changes are correlated with each other.
Statistically, we tested to find that any dependency changes are moderately to strongly correlated to each other for all contribution types, except for the \library~contributions (is-added to is-downgraded, and is-removed to is-downgraded).
We now return to answer RQ1, which we summarize below:

\begin{tcolorbox}[colback=gray!5,colframe=gray!75!black,title= RQ1 Summary]
The alignment between developer contributions and dependency changes varies over time, with the highest values generally observed in earlier stages of the NPM ecosystem. The peaks we identify in the data correspond to a \ecocon~of 0.16, i.e., 16\% of the dependency changes receive aligned contributions. Furthermore, we show that contributions are most congruent with downgrading a dependency, and mainly come from the ecosystem and not the maintainers of either that package or clients.
\end{tcolorbox}

\subsection{Contribution Congruence and Package's Dormancy (RQ2)}

\textit{Approach.}
To address RQ2, we analyze the relationship between the \ecocon~and the likelihood of packages becoming dormant.
In particular, we leverage survival analysis techniques which allow us to model the relationship across different time periods.
We use a Cox proportional-hazard model (i.e., a commonly-used survival analysis model)~\cite{cox_model} to capture the risk of an event (i.e., a package becomes dormant) in relation to factors of interest (e.g., the \ecocon) in the elapsed time.

To do so, we extract the event and the factors for each package in each time period (i.e., a quarter).
In our Cox model, the dependent variable is the event that a package becomes dormant at a particular time point.
In line with prior work~\cite{coelho2017modern, wang2012survival,fse2018_sustained}, a project is regarded as dormant if it is no longer being maintained or does not have development activity.
Specifically, similar to the work of~\cite{fse2018_sustained},
we consider a package as \emph{dormant} in the quarter $q$ if it had an average number of commits less than one for four consecutive quarters (i.e., from the quarter $q$ until the quarter $(q+3)$).
For example, in the first, second, third, and fourth quarters in 2018, a package that has 1, 0, 0, and 1 commits, respectively is considered as \emph{dormant} since the first quarter in 2018 because the average number of commits for four consecutive quarters is 0.5 ($\frac{2}{4}$).
Note that we exclude packages that have the first commit after October 2019, since their historical data is not sufficient to identify whether the package is dormant or not (i.e., less than four consecutive quarters).
As our RQ1 shows that the correlations between different dependency changes for all contribution types vary, it is also possible that these DC congruence may provide various signals to the package dormancy given different dependency changes and contribution types.
Thus, for independent variables in our Cox model, we use 32 metrics which measure the package-level \ecocon~(see Section \ref{sec:data_preparation}).
Since prior work has shown that project characteristics can be associated with the chance of packages becoming dormant~\cite{fse2018_sustained}, we also include six project characteristics (see Table~\ref{tab:project_characteristics}) as control variables in our Cox model.

To construct the Cox model, we use a similar approach as Valiev et al.~\cite{fse2018_sustained}, which has the following four steps.
First, we perform log-transform on the independent variables with skewed distribution to stabilize the variance and reduce heteroscedasticity, which will improve the model fit.
We manually observe the skewness of the distribution using histogram plots.  
Second, we check for the multicolinearity between independent variables using the variance inflation factor (VIF) test.
We remove the variables that have VIF scores above the recommended maximum of 5~\cite{cohen1983applied}.
Third, we test the assumption of constant hazard ratios overtime.
To do so, we employ graphical diagnostics of Schoenfeld residuals~\cite{grambsch1994proportional}.
We observe that Schoenfeld residuals of the remaining variables have a random pattern against time, which implies that the Cox model assumptions are not violated.
Finally, we use \texttt{coxph} in the R survival package to build a Cox proportional-hazards model.

\begin{table}[t]
    \centering
    \caption{Project characteristics~\cite{fse2018_sustained} that are used as control variables in our Cox model.}
    \scalebox{0.75}{
    \begin{tabular}{p{2.5cm}|p{8.7cm}}
    \toprule
         \textbf{Metric} & \textbf{Description}  \\ \hline
         \#Commits & The number of commits occurring in the observed quarter. \\
         \#Contributors & The number of GitHub users who have authored commits in the observed quarter. \\
         Core Team Size & The number of GitHub users responsible for 90\% of contributions in the observed quarter. \\
         \#Issues & The number of issues created in the observed quarter. \\
         \#Non-Developer Issue Reporters & The number of GitHub users who did not author any commits but created an issue in the observed quarter. \\
         License Type & The license type that the studied package used in the observed quarter: Strong copy-left (e.g., GPL, Affero), Weak-copy-left (e.g., LGPL, MPL, OPL), or non-copy-left (e.g., Apache, BSD) \\ 
    \bottomrule
    \end{tabular}
    }
    
    \label{tab:project_characteristics}
\end{table}

To analyze the association between each independent variable and the likelihood of a package becoming dormant, we examine the coefficient values and effect sizes of the independent variables.
The coefficient values in the Cox model are hazard ratios, where a hazard ratio above 1 implies that a variable is positively associated with the event probability, while a hazard ratio below 1 indicates an inverse association. 
For the effect size, we perform ANOVA type-II to obtain Log-likelihood Ratio $\chi^2$ (LR $\chi^2$).
The larger the LR $\chi^2$ of the variable is, the stronger the relation of the variable to the likelihood of a package becoming dormant.

\begin{table}[t]
\caption{Survival analysis statistics}
\label{tab:survival}
\resizebox{.48\textwidth}{!}{
\begin{tabular}{@{}lllrr@{}}
\toprule
\multicolumn{5}{c}{NPM Ecosystem Survival Analysis}                                 \\
\multicolumn{5}{c}{response: dormant = TRUE}                                       \\
\multicolumn{5}{c}{$R^2$ = 30.7\%}                                                 \\ \midrule
\multicolumn{3}{l}{}                                   & Coeffs (Err.) & LR Chisq  \\ \midrule
\multicolumn{3}{l}{Number of commits}                  & 0.98 (\textless{}0.01)$\ast$$\ast$$\ast$   & 5437.5$\ast$$\ast$$\ast$    \\
\multicolumn{3}{l}{Log number of contributors}         & 7.85 (0.01)$\ast$$\ast$$\ast$   & 23699.3$\ast$$\ast$$\ast$   \\
\multicolumn{3}{l}{Log number of non-developer issues} & 0.68 (0.01)$\ast$$\ast$$\ast$   & 4781.0$\ast$$\ast$$\ast$    \\
\multicolumn{3}{l}{Core size of the team}              & 0.14 (0.02)$\ast$$\ast$$\ast$   & 13378.8$\ast$$\ast$$\ast$   \\
\multicolumn{3}{l}{Strong copy-left license (vs none)} & 0.88 (0.02)$\ast$$\ast$$\ast$   & 839.3$\ast$$\ast$$\ast$     \\
\multicolumn{3}{l}{Weak copy-left license (vs none)}   & 0.69 (0.03)$\ast$$\ast$$\ast$   &           \\
\multicolumn{3}{l}{Non-copy-left license (vs none)}    & 0.82 (\textless{}0.01)$\ast$$\ast$$\ast$   &           \\ \midrule
Contribution Type      & Dependency Change  & Type & Coeffs (Err.) & LR Chisq  \\ \midrule
\nonmaintain      & is-added          & issue        & 1.33 (0.32)   & 0.7       \\
\nonmaintain      & is-removed        & issue        & 0.92 (0.15)   & 0.3       \\
\nonmaintain      & is-downgraded     & issue        & 1.26 (0.21)   & 1.1       \\
\textbf{\nonmaintain}     & \textbf{is-upgraded}   & \textbf{issue} & \textbf{0.67 (0.08)}$\ast$$\ast$$\ast$ & \textbf{24.0$\ast$$\ast$$\ast$} \\
\client   & is-added          & issue        & 0.82 (0.18)   & 1.2       \\
\client   & is-removed        & issue        & 0.62 (0.29)   & 3.1       \\
\client   & is-downgraded     & issue        & 0.77 (0.45)   & 0.4       \\
\textbf{\client}  & \textbf{is-upgraded}   & \textbf{issue} & \textbf{0.66 (0.16)}$\ast$$\ast$ & \textbf{7.6$\ast$$\ast$}        \\
\library  & is-added          & issue        & 0.64 (0.27)   & 3.1       \\
\library  & is-removed        & issue        & 1.33 (0.32)   & 0.7       \\
\library  & is-downgraded     & issue        & 0.71 (0.58)   & 0.4       \\
\nonmaintain      & is-added          & PR           & 0.95 (0.14)   & 0.2       \\
\nonmaintain      & is-removed        & PR           & 0.99 (0.18)   & 0.0       \\
\textbf{\nonmaintain}     & \textbf{is-downgraded} & \textbf{PR}    & \textbf{0.51 (0.24)}$\ast$$\ast$ & \textbf{7.1$\ast$$\ast$}        \\
\nonmaintain      & is-upgraded       & PR           & 0.87 (0.09)   & 1.5       \\
\maintain          & is-added          & PR           & 0.87 (0.07)   & 3.9$\ast$ \\
\textbf{\maintain}         & \textbf{is-removed}    & \textbf{PR}    & \textbf{1.31 (0.10)}$\ast$$\ast$$\ast$ & \textbf{6.3$\ast$}              \\
\textbf{\maintain}         & \textbf{is-downgraded} & \textbf{PR}    & \textbf{0.60 (0.19)}$\ast$$\ast$ & \textbf{8.0$\ast$$\ast$}        \\
\maintain          & is-upgraded       & PR           & 0.98 (0.05)   & 0.1       \\
\client   & is-added          & PR           & 1.27 (0.26)   & 0.8       \\
\client   & is-removed        & PR           & 1.27 (0.37)   & 0.4       \\
\client   & is-downgraded     & PR           & 0.79 (0.58)   & 0.2       \\
\client   & is-upgraded       & PR           & 1.08 (0.21)   & 0.1       \\
\library  & is-added          & PR           & 0.97 (0.38)   & 0.0       \\
\library  & is-removed        & PR           & 0.49 (0.51)   & 1.9       \\
\library  & is-downgraded     & PR           & 0.14 (1.37)   & 2.8       \\
\textbf{\library} & \textbf{is-upgraded}   & \textbf{PR}    & \textbf{0.42 (0.20)}$\ast$$\ast$$\ast$ & \textbf{22.0$\ast$$\ast$$\ast$} \\ \bottomrule
\multicolumn{5}{l}{$\ast$ p$<$0.05; $\ast$$\ast$ p$<$0.01; $\ast$$\ast$$\ast$ p$<$0.001}\\
\multicolumn{5}{p{10cm}}{\small{\#Issues, is-added and  \maintain~and  issue, is-downgraded and  \maintain~and  issue, is-removed and  \maintain~and issue, is-upgraded and  \maintain~and  issue, and is-upgraded and  \library~and issue are removed during VIF analysis.}}\\
\end{tabular}
}
\end{table}

Table~\ref{tab:survival} shows that our Cox proportional-hazard model achieves an adjusted $R^2$ of 30.7\%, suggesting that the goodness-of-fit of our model is acceptable as the model is supposed to be explanatory and not for prediction~\cite{r2_small}.

\textit{\textbf{Association between the contribution congruence and the chance of packages becoming dormant.}}
We found that six congruence-related variables are significantly associated  with the likelihood of a package becoming dormant in our Cox model, i.e., \textit{is-removed and \maintain~and PR}, \textit{is-downgraded and \maintain~and PR}, \textit{is-downgraded and \nonmaintain~and PR}, \textit{is-upgraded and \client~and issue}, \textit{is-upgraded and \library~and PR}, and \textit{is-upgraded and \nonmaintain~and issue}.
Among these six variables, \textit{is-upgraded and  \library~and  PR} and \textit{is-upgraded and  \nonmaintain~and  issue} have the largest LR $\chi^2$ values (see Table~\ref{tab:survival}). 
Their coefficient values also indicate an inverse association with the likelihood of a package becoming dormant.
In other words, the coefficient value of \textit{is-upgraded and  \library~and  PR} indicates that the higher the number of \library~PRs that are aligned with a dependency upgrade, the lower the likelihood that a package becomes dormant.
Similarly, the coefficient value of \textit{is-upgraded and  \nonmaintain~and  issue} indicates that the higher the number of \nonmaintain~issues that are aligned with a dependency upgrade, the lower the likelihood that a package becomes dormant. 
The coefficient values of \textit{is-upgraded and  \library~and  PR}, \textit{is-downgraded and  \nonmaintain~and  PR}, and \textit{is-upgraded and  \client~and  issue} also show the same direction of the association.
These results suggest that a package is less likely to become dormant if the contributions (i.e., PRs and issues) are aligned with the events of upgrading or downgrading dependencies.

\begin{tcolorbox}[colback=gray!5,colframe=gray!75!black,title= RQ2 Summary]
Our survival analysis shows that the different types of \ecocon~share an inverse association with the likelihood of a package becoming dormant. 
For instance, the higher the number of issues from specific types (i.e., \nonmaintain, \client) that are aligned with a dependency upgrade, the lower the likelihood that a package becomes dormant.
\end{tcolorbox}

\subsection{Similarities between Contributions (RQ3)}

\textit{Approach.} 
To answer RQ3, we investigate whether the type of aligned contributions submitted to dependencies are the same as other contributions by analyzing the source code similarity and file location similarity.
Hence, we focus on client contributions (i.e., \client), comparing against contributions from the ecosystem  (i.e., \nonmaintain), and then the rest of maintainer contributions (i.e., \maintain~and \library).
We compute similarity based on two common methods to compare the content of PRs, which is (i) source code and (ii) file path similarity \cite{Wang:IST2021, Li2021DetectingDC, JIANG2019196}.
More specifically, we use the source code similarity to measure whether the two contributions share similar source code; and use the file path similarity to measure whether the two contributions modify similar components.
From our sample, we compare the similarities between PRs of the same contribution type of a developer.
As a PR may modify multiple files, we compare all components (i.e., source code lines and file paths).

Our source code similarity measure is based on the Jaccard index of tokens which approximates the edit distance between two PRs~\cite{Ukkonen1992}. 
Following the analysis of repeated bug fixes~\cite{Yue:ICSME2017}, to compare a pair of PRs, we take only added lines by the PRs into account. 
Formally, for any PR ($p$), the source code similarity is defined as follows:
\begin{equation}
\begin{multlined}
    SourceCodeSim(p_1, p_2) = 
    \frac{|trigrams(p_1)\cap trigrams(p_2)|}{|trigrams(p_1) \cup trigrams(p_2)|}
\end{multlined}
\end{equation}
where $trigrams(p)$ is a multiset of trigrams (3-grams) of source code tokens extracted from all source code lines added by a PR ($p$).
A higher similarity indicates that a larger amount of source code is shared.
The similarity is an extension of file similarity defined by Ishio et al.~\cite{Ishio:MSR2017} for PRs.
This measure has been widely adapted in various software engineering studies \cite{Ishio:MSR2017, Jiffriya:2014, Win:2015} and less affected by moved code in a PR, compared with other measures.
On that same note, we apply the file location-based model that is used in typical code review recommendations \cite{Thongtanunam:SANER2015, Yu:IST2016, Wang:IST2021} to compute file path similarity that is defined as follows:
\begin{equation*}
\begin{multlined}
    StringSim_{LCx}(f_1,f_2) = 
    \frac{LCx(f_1,f_2)}{max(Length(f_1), Length(f_2)
)}
\end{multlined}
\end{equation*}

where the $LCx(f_1,f_2)$ function has a parameter to specify how to compare file path components $f_1$ and $ f_2$. 
Four different comparison techniques, i.e., Longest Common Prefix (LCP), Longest Common Suffix (LCS), Longest Common Substring (LCSubstr), and Longest Common Subsequence (LCSubseq) are used in the $LCx$ function.
The comparison function value is normalized by the maximum length of each file path in $f_1$ and $f_2$, i.e., the number of file path components.
To calculate the file path similarity between two PRs, we do a pairwise comparison of all file paths in the two PRs and then summarize these by taking the average score.
Since we use four different comparison techniques, four similarity scores between two PRs are retrieved and we also summarize these by taking the average score.
A higher similarity indicates that a larger amount of components are shared in the PRs.

Once we compute the similarity between PRs in each contribution types (i.e., \client, \nonmaintain, and \maintain~and ~\library), we now analyze whether the similarity of the \client~contributions is different from the similarity of the other contributions.
To visualize this, we will use a box-plot to show the distributions of similarities between PRs in each of the contribution types.

To statistically confirm the differences in the three contribution types, we use the  Kruskal-Wallis H test \cite{Kruskal:1952}. This is a non-parametric statistical test to use when comparing two or more than two types.
We test the null hypothesis that \textit{`two different types have similar contributions.'}
We also measure the effect size using Cliff's $\delta$, a non-parametric effect size measure \cite{Cliff:1993}.
Effect size is analyzed as follows: 
(1) $|\delta| < 0.147$ as Negligible, (2) $0.147 \leq |\delta| <0.33$ as Small, (3) $0.33 \leq |\delta| <0.474$ as Medium, or  (4) $0.474 \leq | \delta|$ as Large.
We use the cliffsDelta\footnote{https://github.com/neilernst/cliffsDelta} package to analyze Cliff's $\delta$.

\begin{figure}[t]
\centering
\includegraphics[width=\linewidth]{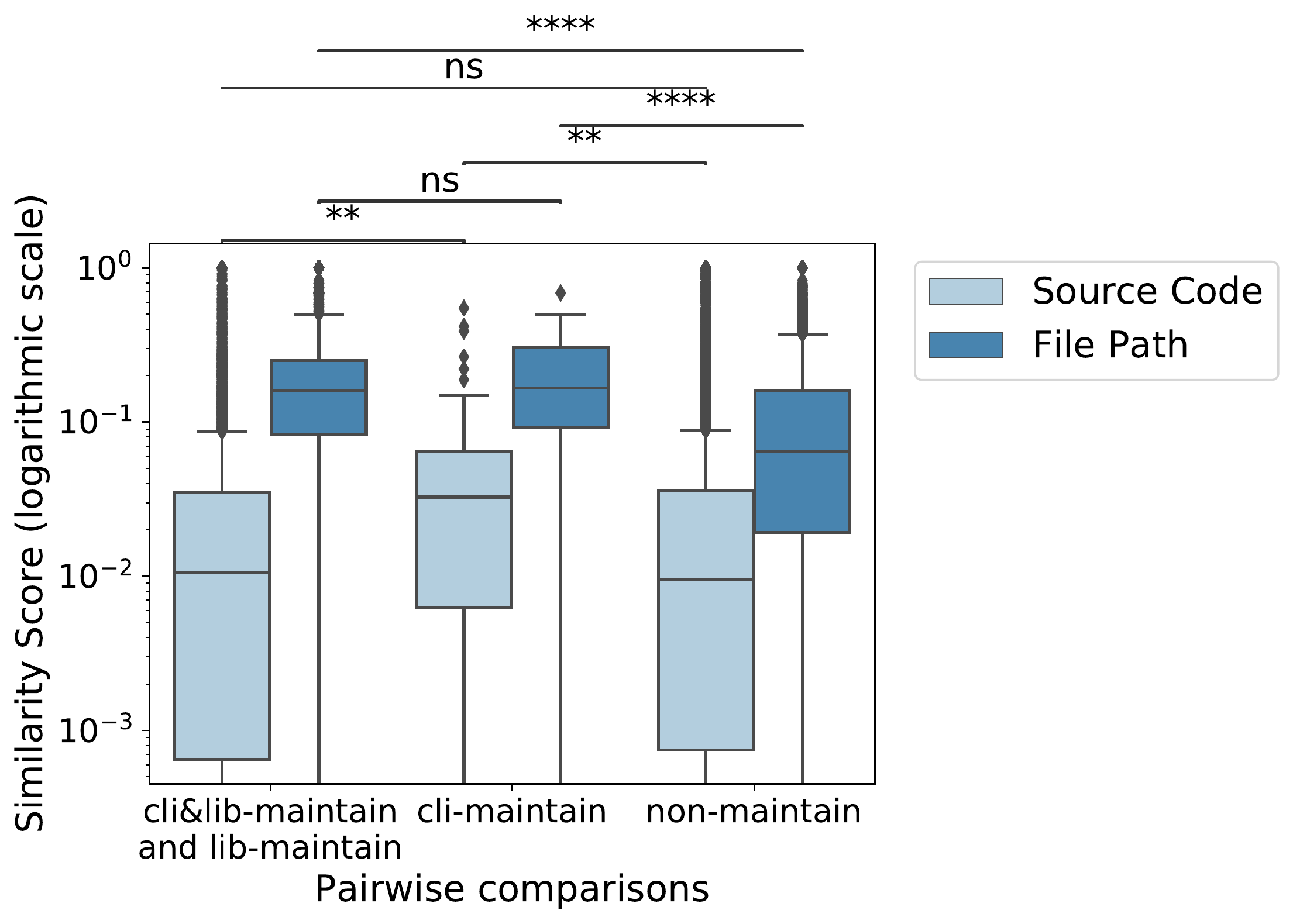}
\caption{Comparisons of contribution types. \footnotesize{(ns: no significance, **: p-value $<$ 0.01, ***: p-value $<$ 0.001, ****: p-value $<$ 0.0001)}}
\label{fig:rq2_result}
\end{figure}

\begin{table}[t]
\centering
\caption{Statistical comparisons of contribution types. }
\label{tab:rq3-stat}
\scalebox{0.85}{
\begin{tabular}{@{}lcc@{}}
\toprule
\textbf{}              & \textbf{Source Code} & \textbf{File Path} \\ \midrule
\client~\textit{vs} (\maintain~and~\library)      & Small                & Negligible         \\
\nonmaintain~\textit{vs} (\maintain~and~\library) & Negligible           & Medium            \\
\client~\textit{vs} \nonmaintain   & Small                & Medium             \\ \bottomrule
\multicolumn{3}{l}{\small{Effect size: Negligible $|\delta| < 0.147$, Small $0.147 \leq |\delta| <0.33$,}}\\
\multicolumn{3}{l}{\small{Medium $0.33 \leq |\delta| <0.474$, Large $0.474 \leq | \delta|$}}\\
\end{tabular}
}
\end{table}

\textit{\textbf{Similarities/differences between client contributions and other contributions.}}
Figure \ref{fig:rq2_result} shows that overall the similarities of contents of PRs in terms of the source code and file paths are relatively low (i.e., scores of 0.01 to 0.17).
This could be explained by our method to aggregate all source code and file paths for comparison, which is different to typical file-level similarity comparison techniques \cite{Ragkhitwetsagul:EMSE2018}.
Relatively, we see that the file path similarity is higher than the source code similarity.

In terms of statistical significance, from Figure \ref{fig:rq2_result}, we make three key observations.
First, file path content between \maintain~and ~\library~and \nonmaintain~contributions are different. 
Second, we find differences in the source code content between \client~and other contributions.
This suggests that the content of aligned contributions submitted to packages that developers depend on may differ from the other contributions made to either their own packages or other packages.
Conversely, we find that there is no significant difference in the source code between \maintain~and \library~and \nonmaintain~contributions, which suggests that developers may be submitting similar code, regardless of whether they are maintaining packages or not.
Confirming statistical significance, we show that the effect size ranges from negligible to medium strength as shown in Table \ref{tab:rq3-stat}.

\begin{tcolorbox}[colback=gray!5,colframe=gray!75!black,title= RQ3 Summary]
Comparing contribution similarity in terms of source code and file paths, we find statistical differences in source code content of aligned contributions submitted to dependencies when compared to those that are not aligned.
In other words, congruent contributions are not typical contributions by that contributor.
\end{tcolorbox}

\section{Discussion}\label{sec:discussion}
We now discuss our results, implications, and threats to validity.

\subsection{Peaks of DC Congruence and Global Events}
The results in RQ1 (i.e., 
Figure~\ref{fig:rq1_result}) show that \ecocon~is changing over time.
Interestingly, we find that the official NPM blog\footnote{Available at \url{https://blog.npmjs.org/}} showed two dependency events that may correlate with the observed peaks.
It is also important to note that we do not aim to draw a causal relationship, but only observe correlations.

\noindent
$\bullet$~\textit{June, 2015 - NPM 3 is released}. According the NPM blog,\footnote{https://blog.npmjs.org/post/122450408965/npm-weekly-20-npm-3-is-here-ish.html} the release has a notable breaking change when migrating to NPM version 3.
According to our \ecocon~values in Figure \ref{fig:rq1_result}, we see the peak changes in the \nonmaintain~contribution and dependency remove congruence (i.e., 0.019 for issues and 0.013 for PRs), and \client~contribution and dependency downgrade congruence (i.e., 0.031 for issues and 0.027 for PRs).

\noindent
$\bullet$~\textit{March 2016 - Left-pad incident}. NPM attracted press attention after a package called \texttt{left-pad}, which many popular JavaScript packages depended on, was unpublished. It caused widespread disruption, leading NPM to change its policies regarding unpublishing packages.  
According to our \ecocon~values in Figure \ref{fig:rq1_result}, we see peak changes in the \nonmaintain~contribution and dependency downgrade congruence (i.e., 0.103 for issues and 0.086 for PRs), and \nonmaintain~contribution and dependency upgrade congruence (i.e., 0.051 for issues and 0.037 for PRs).

\subsection{Implications}\label{sec:results}

\textbf{Package developers: You should welcome contributions from the ecosystem and offer them ways to get involved.}
RQ1 results suggest that contributions that align with dependency changes come from the ecosystem itself, many of them not having any self-interest (i.e., relying on these packages).
The results of our survival analysis in RQ2 show that a package that received contributions aligned with the dependency changes is less likely to become dormant, which implies that packages are indeed dependent on the ecosystem support.
RQ3 shows that \client~contributions differ in source code content when compared to developers' typical contributions. 
These findings underline the important role of the ecosystem.

\textbf{Package users: You might have to contribute to packages you are using to reduce the risks of packages becoming dormant (in multiple possible ways).}
In our RQ2, we found that a large number of packages become dormant in the early stages, with more than 90\% of packages becoming inactive until the end-time of our study.
Taking into account that contributions aligned with dependencies share a relationship with the likelihood of a package becoming dormant, we encourage package users to take any opportunity to contribute back to packages that they depend on.

Currently there is numerous tools to assist with awareness of when a new version of an existing dependency is available (bots like dependabot\footnote{\url{https://github.com/dependabot}}), however, making users aware of raised issues, and potential threats (such as vulnerabilities) to their dependency may cause these users to assist.
Furthermore, the results of RQ3 show that aligned contributions submitted to dependencies are different in source code, which confirms our hypothesis and indicates that package users might be willing to go out of their way to assist with the dependency that they rely on.

\textbf{Researchers: Ecosystem-level is important, do not put all contributions into the same bucket when analyzing open source contributions.} 
Our results highlight the important relationship between dependencies in an ecosystem and contributions. Researchers interested in studying characteristics of open-source developers, e.g., motivation~\cite{gerosa2021shifting} or success~\cite{trinkenreich2021pot}, should consider whether or not the contribution is from the ecosystem. This also applies to researchers working on helping newcomers make their first contribution in open source projects. Several approaches have been proposed to help newcomers find good first issues, e.g., Tan et al.~\cite{good_first_issue}. These approaches generally ignore the relationship of a newcomer to the repository that contains the issues.

\textbf{Ecosystem Governance: 
We suggest that policies and good governance is needed to help ecosystem members.} 
Our results suggest that contributions congruent with dependency changes are from other packages that belong to the same ecosystem, suggesting that it may impact on ecosystem health, resilience, and governance. 
This points to implications such as implementing collective support into existing policies and governance.
Examples include ecosystem guidelines for maintainers and users\footnote{NPM ecosystem governance \url{https://github.com/nodejs/package-maintenance/blob/main/Governance.md}}.
Particularly, RQ2 results indicated that a package is less likely to become dormant if contributions are aligned with dependencies.
%and RQ3 results showed that the source code content of aligned contributions submitted to dependencies is different from developers' typical contributions.
One example of challenges in the relationship between ecosystem members that depend on each other is illustrated by the following quote: `... maintainers have been working sleeplessly on mitigation measures; fixes, docs, CVE, replies to inquiries, etc. Yet nothing is stopping people to bash us, for work we aren't paid for, for a feature we all dislike yet needed to keep due to backward compatibility concerns ...'\footnote{\url{https://twitter.com/yazicivo/status/1469349956880408583?lang=en}}
One strategy to prevent member packages from becoming dormant, is to promote explicit best practices and some ecosystem-wide policies that encourage members contribute to each other.
One example is how the NPM ecosystem has a global blog and central vulnerability advisory\footnote{https://docs.github.com/en/code-security/supply-chain-security/managing-vulnerabilities-in-your-projects-dependencies/browsing-security-vulnerabilities-in-the-github-advisory-database}, to provide a global awareness of ecosystem threats.

\subsection{Threats to Validity}
\label{sec:ttv}

\textbf{Internal Validity} - We discuss five threats to internal validity. The first threat is the correctness of techniques used in our mining task.
We use the listed dependencies and version numbers
as defined in the package.json meta-file. 
Since we based our mining techniques on prior work, we are confident that our results are replicable.
The second threat is related to the constructed matrices for measuring the DC congruence.
Since we pioneer the investigation of DC congruence, we constructed binary matrices in this study.
We will calculate weight matrices, e.g., a matrix captures the number of contributions (issues and PRs) to investigate the impacts of contributions on the DC congruence, in future work.
The third threat is related to the results derived from the survival model.
Although we observe association between explanatory and dependent variables, we cannot infer causal effect of explanatory variables.
Thus, future in-depth qualitative analysis or experimental studies are needed.
The fourth threat is related to the factors used in our survival model.
Other factors might also influence the chance of packages becoming dormant. Since our factors are based on Valiev et al.~\cite{fse2018_sustained}, we are confident in the factors used in our study.
Finally, tool selection has its threats, as different tools and techniques may lead to different results (e.g., Jaccard index for the source code similarity calculation and the statistical testing). 
Our approach involves the use of Pydriller and GitHub Rest API.
We are confident in our results, since all employed tools and techniques are those that have been used in prior work and are well-known in the software engineering community.

\textbf{External Validity} - The main threat to external validity is the generalizability of our results to other package ecosystems. 
Since the NPM ecosystem is the largest package ecosystem, we believe that our model and measurements could easily be applied to other ecosystems that have similar package management systems, such as PyPI for Python and Maven for Java. We will explore this in future work.

%% file: sections/5related_work.tex
\section{Related work}\label{sec:related_work}
\textbf{Socio-Techical Congruence (STC).}
Cataldo et al. \cite{10.1145/1180875.1180929} pioneered the concept of socio-technical congruence (STC), the match between task dependencies among people and coordination activities performed by individuals.
Their follow-up studies (Cataldo et al. \cite{Cataldo:2008ESEM}, Cataldo and Herbsleb \cite{6205767}) then investigated software quality and development productivity.
Further studies also presented STC measures from different perspectives. 
For instance, Valetto et al. \cite{4228662} proposed a measure of STC in the view of network analysis, while Kwan et al. \cite{kwan2009weighted} measured weight STC and the impact of STC on software build success (Kwan et al. \cite{kwan2011}).
Additionally,  Wagstrom et al. \cite{doi:10.5465/ambpp.2010.54500789} measured individualized STC.
Other studies then measured STC from other perspectives such as dependencies, knowledge, and resources \cite{Jiang2012},  global software development \cite{portillo2013agent}, and for file-level analysis \cite{ZHANG201921} and \cite{Mauerer2019}.
\textit{Different from these studies,} we formulate the time-based ecosystem-level \ecocon, adapting the STC measure proposed by Cataldo et al. \cite{10.1145/1180875.1180929}.
Further, we analyze the time-based package-level  \ecocon, making an adjustment on the STC measure proposed by Wagstrom et al. \cite{doi:10.5465/ambpp.2010.54500789}.

\textbf{Package Dependency Changes.}
The research community has carried out a large body of research on understanding the large networks of software package dependencies.
Some studies focused on a single ecosystem, focusing on the updating and lags within that ecosystem.
For instance, Wittern et al~\cite{wittern2016look} analyzed a subset of NPM packages, focusing on the evolution of characteristics such as their dependencies and
update frequency. 
Abdalkareem et al.~\cite{10.1145/3106237.3106267} also empirically analyzed ``trivial'' NPM packages, to suggest that depending on trivial packages can be useful and risk-free if they are well implemented and tested.
Zerouali et al .~\cite{2018_technique_lag} analyzed the package update practices and technical lag to show a strong presence of technical lag caused by the specific use of dependency constraints.
Recent studies include understanding the lag when updating security vulnerabilities \cite{chinthanet2021lags} and dependency downgrades \cite{Cogo:TSE2019} for the NPM ecosystem.
There are works that studied other ecosystems.
Robbes et al.~\cite{Robbes:fse2012} showed that a number of API changes caused by deprecation can have a large impact on the Pharo ecosystem.
Blincoe et al.~\cite{blincoe2019reference} proposed a new method for detecting technical
dependencies between projects on GitHub and IBM.
Bavota et al.~\cite{Bavota:emse2015} studied Apache projects to discover that a client project tends to upgrade a dependency only when substantial changes (e.g., new features) are available.
Kula et al .~\cite{kula2018developers} revealed that affected developers are not likely to respond to a security advisory for Maven libraries, while Wang et al.~\cite{Wang2020AnES} revealed that developers need convincing and fine-grained information for the migration decision.
Most recently, He et al.~\cite{2021_minhui_library} proposed a framework for quantifying dependency changes and library migrations based on version control data using Java projects. 

Other studies focused on analyzing multiple ecosystems, comparing the different attributes between them. 
Decan et al .~\cite{decan2017empirical} showed the comparison of dependency evolution and issue from three different ecosystems.
Similarily,
Kikas et al.~\cite{10.1109/MSR.2017.55}, studied the dependency network structure and evolution of the JavaScript, Ruby, and Rust ecosystems.
Dietrich et al.~\cite{dietrich2019dependency} studied how developers declare dependencies across 17 different package managers.
Stringer et al.~\cite{stringer2020technical} analyzed technical lag across 14 package managers and discovered that technical lag is prevalent across package managers but preventable by using semantic versioning based declaration ranges.
\textit{Differently, in our work,} we focus on how developer contributions align with dependency changes at the ecosystem level, our motivation is different, as we explore how dependency changes correlate with a package becoming dormant.

\textbf{Sustainability of Software Ecosystem}
As with prior work~\cite{2017_tom}, we view software ecosystems as socio-technical networks consisting of technical components (software packages) and social components (communities of developers) that maintain the technical components.
Sustainability is particular visible in the complex and often brittle dependency chains in OSS package ecosystems like NPM, PyPi, and CRAN~\cite{emse2019_Decan}.
Other studies have studied sustainability for R
\cite{2013_r}, Ruby
\cite{ruby2017}, Python
\cite{fse2018_sustained}, and NPM
\cite{2019_promise_mockus}.
Through a longitudinal empirical study of Cargo’s dependency network, Golzadeh~\cite{Golzadeh:2019} revealed that dependencies to a project lead to active involvement in that project.
However, the effect of \ecocon~from different types of contributions on the ecosystem sustainability is still unknown.
\textit{Different to these studies,} we take a quantitative approach to investigate the relationship between the \ecocon~and the likelihood of packages becoming dormant.

\section{Conclusion}\label{sec:conclusion}
This work investigates how developers whose code depends on other packages in the NPM ecosystem, submit contributions to assist these packages.
Borrowing socio-technical techniques, we were able to measure the congruence between these aligned contributions, and changes in dependencies for the ecosystem.
Results indicate that packages indeed rely on the ecosystem for contributions, and that these aligned contributions share a relationship with the likelihood of a package becoming dormant, suggesting that the alignment between contributions and dependency changes could be used as an indicator of package dormancy.
Moreover, the aligned contributions submitted to dependencies differ from developers' typical contributions. 
Our results show that developers do depend on the ecosystem, and should be encouraged to support packages that they depend on. 

\section*{Acknowledgement}
This work is supported by Japanese Society for the Promotion of Science (JSPS) KAKENHI Grant Numbers 20K19774 and 20H05706.
P. Thongtanunam was partially supported by the Australian Research Council’s Discovery Early Career Researcher Award (DECRA) funding scheme (DE210101091).